# Blister Test to Measure the Out-of-Plane Shear Modulus of Few-Layer Graphene

Metehan Calis[1], Narasimha Boddeti[2], and J. Scott Bunch[1,3*]

[1]Boston University, Department of Mechanical Engineering, Boston, MA 02215 USA

[2]Washington State University, School of Mechanical and Materials Engineering, Pullman, WA 99163 USA

[3]Boston University, Division of Materials Science and Engineering, Brookline, MA 02446 USA

*e-mail: bunch@bu.edu

## Abstract

We measure the out-of-plane shear modulus of few-layer graphene (FLG) by a blister test. During the test, we employed a monolayer molybdenum disulfide ($MoS_2$) membrane stacked onto FLG wells to facilitate the separation of FLG from the silicon oxide ($SiO_x$) substrate. Using the deflection profile of the blister, we determine an average shear modulus $G$ of $0.97 \pm 0.15$ GPa, and a free energy model incorporating the interfacial shear force is developed to calculate the adhesion energy between FLG and $SiO_x$ substrate. The experimental protocol can be extended to other two-dimensional (2D) materials and layered structures (LS) made from other materials ($WS_2$, hBN, etc.) to characterize their interlayer interactions. These results provide valuable insight into the mechanics of 2D nano devices which is important in designing more complex flexible electronic devices and nanoelectromechanical systems.

KEYWORDS: Two-dimensional materials, MoS$_2$ on few-layer graphene, shear modulus, layered structures, blister test

## 1. Introduction

Atomically thin graphene, known for its high elastic modulus (Young's modulus ~1TPa)[1], extreme bendability[2], and conformity[3] to a surface, is a great candidate for flexible electronics[4] and soft robotics applications[5] as it can bend and shape into complex structures. Nevertheless, while research on 2D materials and their LS has predominantly concentrated on electrical and optical characterization, there has been a lack of focus on the characterization of the mechanical properties and interactions between the layers of these structures. Moreover, as the thickness reduces to the atomic level, surface forces play a critical role in device fabrication and functionality. For instance, the majority of fabrication processes entail transferring 2D materials from one substrate to another by utilizing van der Waals transfer methods.[6,7] Therefore, a comprehensive understanding of the mechanical properties of 2D materials, both in monolayer and layered configurations, is critical.

There is a growing body of literature demonstrating the strong adhesion energies between 2D materials and various substrates along with the high elastic moduli of these materials. However, one less studied but nevertheless important elastic constant is the shear modulus. The shear modulus which influences the bending rigidity and flexibility is of great importance to van der Waals bonded 2D materials and LSs. The shear modulus plays a role in how the structure folds[8], ripples[9], and slips[10]. Therefore, determining the shear modulus precisely has critical importance.[11] While much is understood about the shear modulus in bulk form of 2D materials such as graphite, less is known about the shear modulus in thin 2D materials and their LSs made from 2D materials. The majority of existing relevant research is based on computational modeling[12–15] and there is only a limited number of experimental studies[8,16–18] focused on the out-of-plane shear modulus of few-layer 2D systems.

## 2. Materials and methods

### 2.1. Materials and measurements

Here, we introduce a new approach to determining the out-of-plane shear modulus of an FLG by using a constant-N pressurized blister test which has been widely used to determine several mechanical properties of 2D thin-films including Young's modulus[19], adhesion energies[20,21], coefficient of friction[22], and shear stress.[10] To fabricate the devices used in the blister test, we start with mechanical exfoliation of FLG over a $SiO_x$/Si wafer. Subsequently, we etch microcavities through the FLG, and $SiO_x$, and into the Si substrate (see the supporting information for details). Then, we use an optical microscope[23] to identify FLG flakes on the substrates that fall within 10 to 50 layers of graphene thickness. Subsequently, we verify their thickness with the atomic force microscope (AFM).[24] Next, utilizing a micro-manipulator, a chemical vapor deposition (CVD)-grown monolayer $MoS_2$ flake is transferred over the wells to cover the microcavities that are etched through the exfoliated FLG. This transferred $MoS_2$ flake enables us to lift and delaminate the FLG at the end of the blister test (Figure 1a) (see the supporting information). After fabrication, the devices are placed into a pressure chamber and charged to a certain input pressure ($p_0$) with argon gas. The gas molecules diffuse into the sealed microcavity and we wait (~ 48 hours) until the input pressure and internal pressure ($p_{int}$) equilibrate ($p_0 = p_{int}$).[25] After taking the devices out of the pressure chamber, $p_{int}$ is greater than $p_{ext}$ ($p_{ext} \equiv p_{atm} \approx 1$ atm), and the $MoS_2$ membrane bulges upward which we image using the AFM. The devices are then returned to the pressure chamber at a higher $p_0$ and this process is repeated at this new input pressure (Figure 1b left). Initially, the bulge radius is equal to the well radius ($a_0$) and the deflected MoS$_2$ behaves as a pressurized circular membrane clamped along the well boundary[25] by the adhesive forces between the MoS$_2$ membrane and the FLG. We use Hencky's model to determine the two-dimensional Young's modulus ($E_{2D}$)[3,19,20,24,26] of MoS$_2$ (see the supporting information for details). Beyond a critical pressure, the pressure load on the MoS$_2$/FLG LS is large enough to overcome the adhesive forces clamping the MoS$_2$/FLG LS to the SiO$_x$ surface, and it delaminates from the surface (Figure 1b. right). In Figure 1c, we show an AFM image of the device before (top) and after delamination (bottom). In Figure 1d, we show cross-sections of the AFM scans of a device, which pass through the center of the blister, corresponding to varying input pressures. Some devices undergo multiple delaminations to larger radii ($a$) at higher pressures. We observe that the MoS$_2$/FLG LS delaminates from the substrate instead of just the MoS$_2$ membrane suggesting that the work of separation between MoS$_2$ and FLG is larger than that of FLG and SiO$_x$. In this delamination, unlike the typical blister configuration, we also observe a kink in the delamination profile where monolayer MoS$_2$ meets the FLG (Figure 1d). This suggests that the FLG layer beneath the MoS2 membrane has separated. We assume that all layers of the FLG delaminate with the MoS$_2$. However, there remains the possibility that some layers of FLG remain attached to the SiO$_x$ surface.

## 2.2 Theoretical Model

We model each MoS$_2$/FLG LS device as a thermodynamic system which includes the membrane, MoS$_2$/FLG LS- substrate interface, trapped gas, and external atmosphere. Our aim is to minimize the free energy of this system to determine its equilibrium configuration at any prescribed input pressure. We built our model based on previous studies[3,27,28], and the free energy of the system can be expressed as:

$$F = F_{mem} + F_{gas} + F_{ext} + F_{adh} \tag{1}$$

$F_{mem}$ is the strain energy of the membrane due to the pressure load assuming axisymmetric deformation, $F_{gas}$ is the energy change due to the expansion of the gas molecules trapped in the blister, $F_{ext}$ is the energy change of the external environment, and $F_{adh}$ is the adhesion energy of the LS - substrate interface (see the supporting information for further details). We incorporate the following assumptions into our strain energy calculations: (1) stretching in the FLG layers is negligible[29] and the MoS$_2$/FLG LS experiences only shear deformation[30], (2) the contribution of bending energy is neglected, and (3) a clamped boundary condition is valid.[19,25] We neglect the bending strain energy contribution to the free energy as it is negligible compared to the shear strain energy (see the supporting information (section 7) for more information). Thus,

$$F_{mem} = \int_0^a \left(\frac{1}{2}(N_r \epsilon_r + N_t \epsilon_t)\right) 2\pi r dr + \int_{a_0}^a \frac{1}{2} G_{2D} \left(\frac{dw}{dr}\right)^2 2\pi r dr \tag{2}$$

where $r$ is the radial coordinate, and $w$ is transverse deflection. $N_r$ and $\varepsilon_r$ are the radial stress and strain, and $N_t$ and $\varepsilon_t$ are the circumferential stress and strain. $G_{2D}$ is the two-dimensional shear modulus of FLG. We assume that the contribution of the MoS$_2$/FLG interface to $G_{2D}$ is negligible, as the LS is largely composed of graphene layers. Therefore, $G_{2D}$ can be calculated by multiplying the shear modulus ($G$) by the thickness of the layered structure ($G$ x *(thickness of LS)* = $G_{2D}$ *(N/m)*). We modified Williams' model[31] for pressure-loaded clamped axisymmetric membranes by adding a shear term to the force balance equation. This allows us to determine the $G_{2D}$ of the FLG when $E_{2D}$ is known. In this model, we sub-divide the whole blister into two regions: (i) Region I ($r \leq a_0$) where only MoS$_2$ is suspended, and (ii) Region II ($a_0 < r \leq a$) comprised of the delaminated MoS$_2$/FLG LS. The deflection profiles, denoted as *w(r)*, of the delaminated LS device can be expressed as follows:

$$w(r) = \begin{cases} w_1(r) = \left(\frac{pa^4}{Et}\right)^{\frac{1}{3}} \sum_{j=0,1,\ldots} C_j \zeta^{2j}, & Region\ I\ (r \leq a_0) \\ w_2(r) = \left(\frac{pa^4}{Et}\right)^{\frac{1}{3}} \sum_{j=0,1,\ldots} B_j \left(1 - \zeta^{(2j+2)}\right), & Region\ II\ (a_0 < r \leq a) \end{cases} \quad (3)$$

where $p$ is the pressure difference across the membrane ($p = p_{int} - p_{ext}$), $t$ is the thickness of the membrane, and $\zeta = \frac{r}{a}$. The coefficients $C_j$ and $B_j$ are functions of $f_0 = 4G_{2D}/(E_{2D}p^2a^2)^{1/3}$ and are determined utilizing the governing equilibrium equations, clamped boundary conditions, and continuity of the displacements (see the supporting information for further details). We take $E_{2D}$ of the MoS$_2$ layer ($E_{2D} = E_{bulk}$ x (MoS$_2$ thickness)) as Young's modulus of the whole system since only the MoS$_2$ membrane is assumed to stretch by the pressure load. With $f_0$ as the fitting parameter, we fit the deflection profile from our model (dashed line in Figure 1d) to the AFM cross-section of the delaminated blister configuration to determine $f_0$. Assuming isothermal expansion of a fixed number of gas molecules, the ideal gas law can be written as $p_0V_0 = p_{int}(V_0+V_b)$ where $V_0$ is the initial volume of the microcavity. With the best-fit profile, we calculate the bulge volume ($V_b$) which allows us to determine the pressure difference $p$ and thus, $G_{2D}$. Next, we incorporate the calculated $G_{2D}$ value into the free energy model to find the adhesion energy between FLG and the SiO$_x$ substrate when the blister reaches its equilibrium delaminated configuration.

### 3. Result and discussion

We measure the blister profile of each device that shows MoS$_2$/FLG LS delamination from the substrate, and in Figure 2 we plot the shear modulus for 10 devices with nine of them delaminating multiple times (2 or 3 times). Additionally, any irregular and non-circular LS delaminations were not included in the analysis (see the supporting information). We find the average shear modulus for FLG to be $G = 0.97 \pm 0.15$ GPa. Our finding suggests that the primary contribution to the shear modulus of the LS is the FLG since the primary component of our experimental devices consists of graphite (FLG thickness range ≈ 4.25 nm - 6.25 nm versus monolayer MoS$_2$ thickness = 0.65 nm).[32] This value aligns with previous experimental studies on the out-of-plane shear modulus of various types of graphite (0.36 – 4.52 GPa) and is close to the value of G = 4 GPa for intrinsic dislocation-free graphite.[8,16–18]

The only unknown parameter remaining in the description of the thermodynamic system is the separation energy ($\Gamma_{sep}$) of the FLG layers from the SiO$_x$ surface. Through an iterative process, we numerically determine the separation energy for each delaminated device by matching the theoretical system minimum with respect to $a$ and $p$ with the experimental observation. In Figure 3a, we plot the free energy change of one device as a function of $a$ and $p$. The red dot on the plot shows the free energy minimum that matches the experimental observation and thus provides us the separation energy. The separation energies thus obtained are shown in Figure 3b, with the average $\Gamma_{sep}$ = 0.2 ± 0.02 J/m$^2$ shown as a dashed line. Our measured value for the separation energy aligns with previous studies[19,21,33] on the separation of multilayer graphene over SiO$_x$ substrates. Our model can be extended to predict the delamination behavior of other LS devices that are made of other 2D materials (see the supporting information).

To explore the impact of the thickness of the FLG on the delamination behavior, we transfer monolayer MoS$_2$ onto FLG flakes with varying thicknesses, targeting a total LS thickness range of 2 nm to 11 nm since after a certain thickness of FLG, only MoS$_2$ membrane delaminates from the surface.[25] In Figure 4, we show the optical microscope image along with thickness measurements and blister cross-sections. We include an example of a thinner device (2 nm) as shown in Figure 4a where the kink on the blister profile of the LS delamination is not noticeable in the AFM scan (see the supporting information for further details). As the underlying graphite thickness in the LS approaches the thickness of the monolayer MoS$_2$ membrane, it is not possible to determine whether LS delamination has occurred or not since the layers of the delaminated LS blisters become more flexible and compatible with each other (see Figure S8). Consequently, we are unable to precisely fit the curve to the cross-section of the delaminated device, which makes the determination of $G_{2D}$ for such devices difficult. Figure 4b is an explanatory example in which we observe both only MoS$_2$ and MoS$_2$/FLG LS delaminations in two different devices at the same input pressure on the same multilayer graphene flake which is 6.3 nm thick. The thickness of all the devices utilized for the calculation of $G_{2D}$ lay within the range of 5 - 7 nm. In contrast, as shown in Figure 4c, above a certain graphite thickness, we observe only delamination of the MoS$_2$ membrane from the FLG surface. Additionally, as shown in Figure 1c, thin MoS$_2$/FLG devices are prone to developing wrinkles which can affect the assumption of a fully clamped, axisymmetric profile. We neglect the effect of wrinkles because there is no simple analytical model available to accurately describe their influence.

To further understand why we do not observe LS delamination for thicknesses exceeding a value of ~7 nm, we will compare the free energy variation at three different input pressures. In Figure 5, we plot the variation in free energy according to: (i) the standard free energy model based on Hencky's solution that describes the delamination of the MoS$_2$ membrane from FLG (solid lines), and (ii) the LS free energy model that we utilize to describe the delamination of the MoS$_2$/FLG LS (dashed lines), both expressed as a function of the blister radius. In this demonstration, we focus on three different thicknesses of LS, 2 nm, 6 nm, and 9 nm. For each case, we use microcavity dimensions that represent the devices used in the experiment: depth of 600 nm and a radius of 2.5 µm and use the experimentally measured parameters $G = 0.97 \pm 0.15$

GPa, $\Gamma_{sep}$ = 0.2 J/m² (work of separation of graphite from SiO$_x$), $E_{2D}$ =171.1 N/m, and $\Gamma_{sep}$ = 0.39 J/m² (work of separation of MoS$_2$ from graphite) from our previous work.[25] For each free energy model, we locate the equilibrium configuration by finding the local minimum of the free energy functions ($F$) by setting its derivatives with respect to the independent variable $a$ to zero ($dF/da$ =0). When the input pressure $p_o$ is below a critical pressure ($p_{cr}$) for delamination specific to the free energy model, the membrane stays pinned at the initial radius since there are no local minima above the well radius. When $p_o = p_{cr}$, the system possesses an equilibrium configuration at $a = a_0$ and further input pressure increase beyond this point results in delamination to $a > a_0$ $F$ reaches its local minimum (the part of the curve with $a < a_0$ is not observable physically).

In Figure 5a (LS thickness: 2 nm), we can see that the LS model possesses a local minimum at the edge of the well at an input pressure lower than the standard blister model. Therefore beyond this critical pressure $p_{cr}$, for the LS model, the system will follow the minimum of the LS model and LS delamination is expected. In Figure 5b (LS thickness: 6 nm), the local minima for both free energy models occur around the same critical input pressure. Thus variation in the adhesion strength or the local van der Waals interactions play a crucial role in determining whether the system follows one path or the other for the delamination configuration (see the supporting information for details on identifying the delamination transition zone). As FLG thickness increases, MoS$_2$ membrane separation from the FLG surface becomes thermodynamically more favorable than FLG delamination from the SiO$_x$ substrate and FLG shearing. In Figure 5c, with thicker FLG layer combinations (LS thickness: 9 nm), the standard model reaches a local minimum before the LS model. So only MoS$_2$ separates from the graphite surface and no LS delamination takes place.

## 4. Conclusion

We conducted a study on the mechanical behavior of a FLG using the constant-N blister test. By increasing the input pressure, causing the $MoS_2$ membrane to bulge upwards, we successfully induced delamination of the $MoS_2$/FLG LS from the $SiO_x$ surface. Analyzing the blister configuration as a thermodynamic system, we measure the shear modulus of the FLG as $G = 0.97 \pm 0.15$ GPa and $\Gamma_{sep} = 0.2 \pm 0.02$ J/m$^2$ for separation energy for FLG from $SiO_x$ surface. We also calculated the critical pressure and thickness relation to demonstrate that beyond ~7 nm thickness of $MoS_2$/FLG, we do not observe LS delamination. Our study is useful for understanding the mechanical behavior of layered 2D structures and can be extended to determine the shear modulus of 2D layered heterostructures. This can be used to guide the development of new designs of electrical and mechanical systems in flexible electronic[34] and soft robotics applications[35] or the fabrication of more complex structures based on 2D heterostructure materials.[36,37]

**Conflicts of interest**

There are no conflicts to declare.

**Supporting information**

Supporting information covers the CVD growth and characterization, FLG well fabrication and MoS$_2$ layers transfer procedures, Young's modulus calculation, shear modulus derivation, and free energy model, investigation of the delamination behavior of the LS System by varying the thickness, Raman spectroscopy analysis over the delaminated blister, bending modulus and bending strain energy calculation, thickness dependence of delaminated LS profile, non-circular delamination examples.

**Acknowledgments**

This work was funded by the Ministry of National Education of Turkey under the Graduate Education Scholarship (YLSY) program.

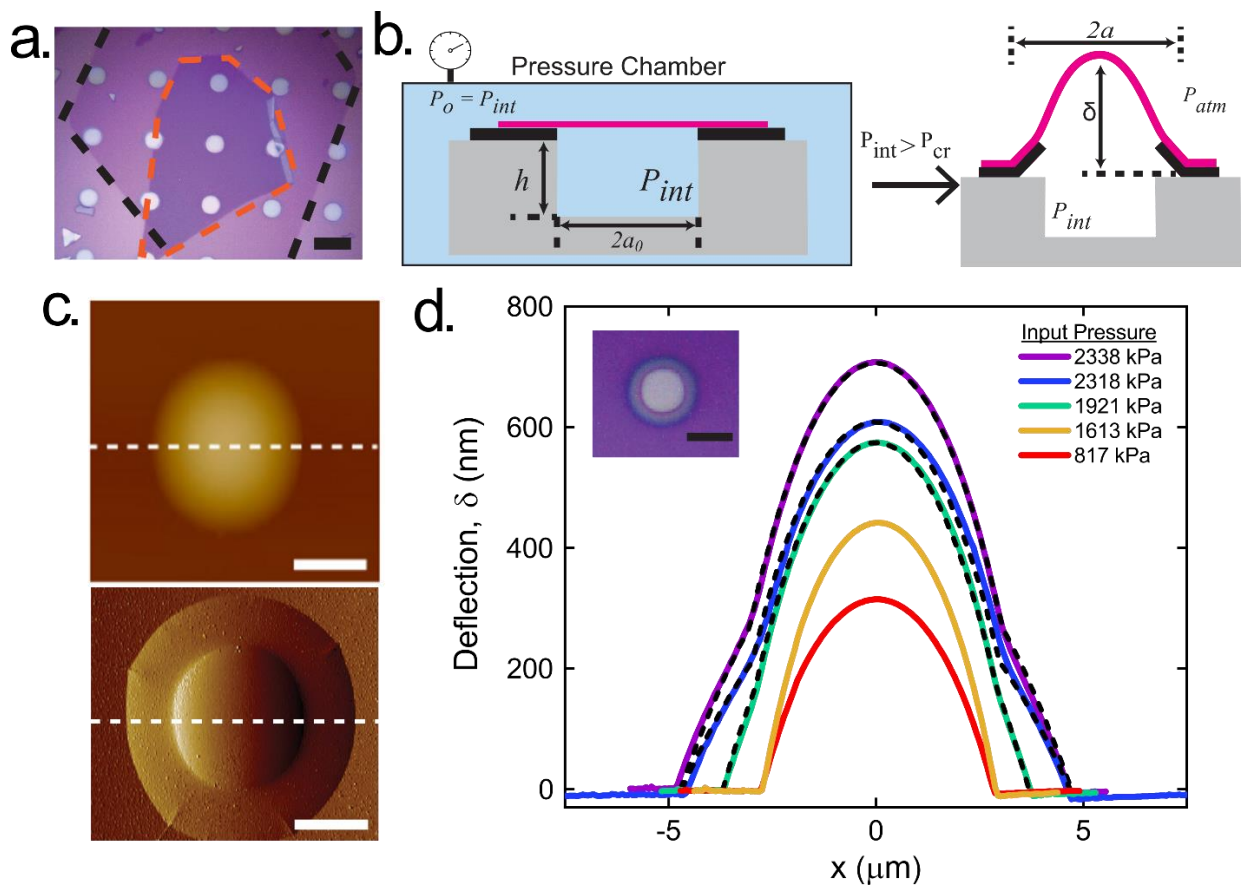

**Figure 1. a)** Optical image of the monolayer MoS$_2$ over the FLG substrate. The orange and black dashed lines show the boundary of monolayer MoS$_2$ flakes and FLG substrate, respectively. The scale bar is 10 μm. **b)** Schematic illustration of the experimental procedure. Devices are kept in the pressure chamber until $p_0 = p_{int}$ (left). When the devices are taken out, the MoS$_2$ membrane bulges up due to $p_{int} > p_{ext}$ ($\approx p_{atm}$). This process is repeated with higher input pressures until the LS delamination is observed from the SiO$_x$ surface ($p_{int} > p_{crt}$) (right). (pink: monolayer MoS$_2$ membrane, black: FLG, grey: SiO$_x$/Si substrate) **c)** AFM height image of the MoS$_2$/FLG LS devices before (up) and AFM amplitude image after delamination (down). Scale bars are 2.5 μm. **d)** Representative AFM cross-sections of the devices at various input pressures. In this particular device, we observed 3 LS delaminations. The dashed line curves are deflection profile fittings for the delaminated configurations, used to calculate the shear modulus. The kinks at the blisters become more pronounced as the input pressure increases. (Inset: Optical image of the delaminated MoS$_2$/FLG LS device. Scale bar is 5 μm.)

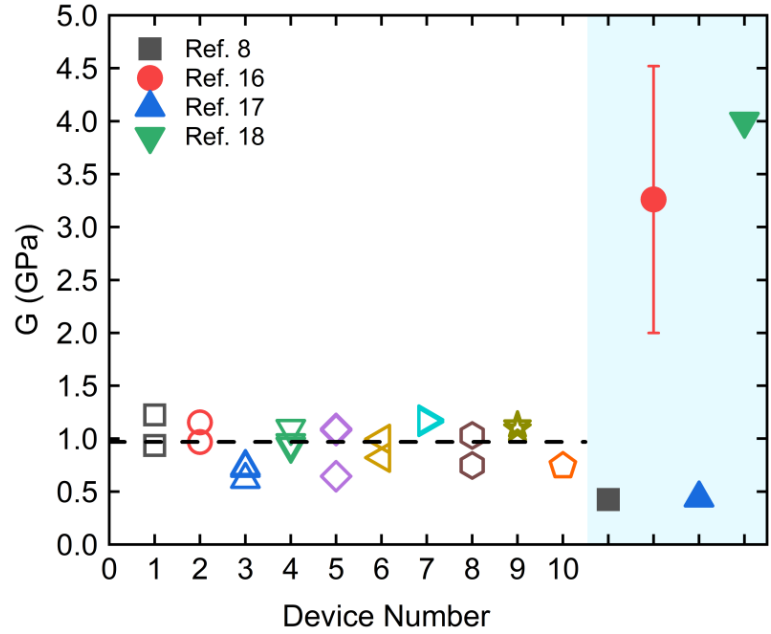

**Figure 2.** Shear modulus of the 10 devices. 9 devices undergo multiple delaminations from the surface. The dashed line is the average of all devices ($G = 0.97 \pm 0.15$ GPa). The data in the blue shaded box are taken from the literature.

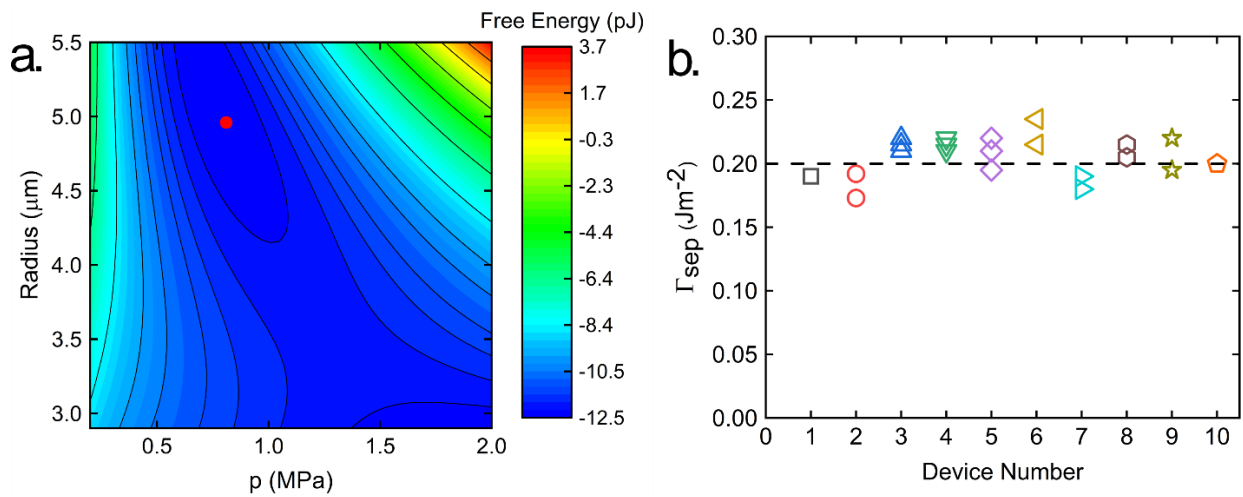

**Figure 3. a)** Filled contour plot of free energy change of MoS$_2$/FLG LS. The red dot in the plot shows the location of the minimum energy after delamination. **b)** Separation energies of the MoS$_2$/FLG LS devices from SiO$_x$. Several devices were subjected to multiple delamination. The dashed line is the average of all devices ($\Gamma_{sep}$ = 0.2 ± 0.02 J/m$^2$).

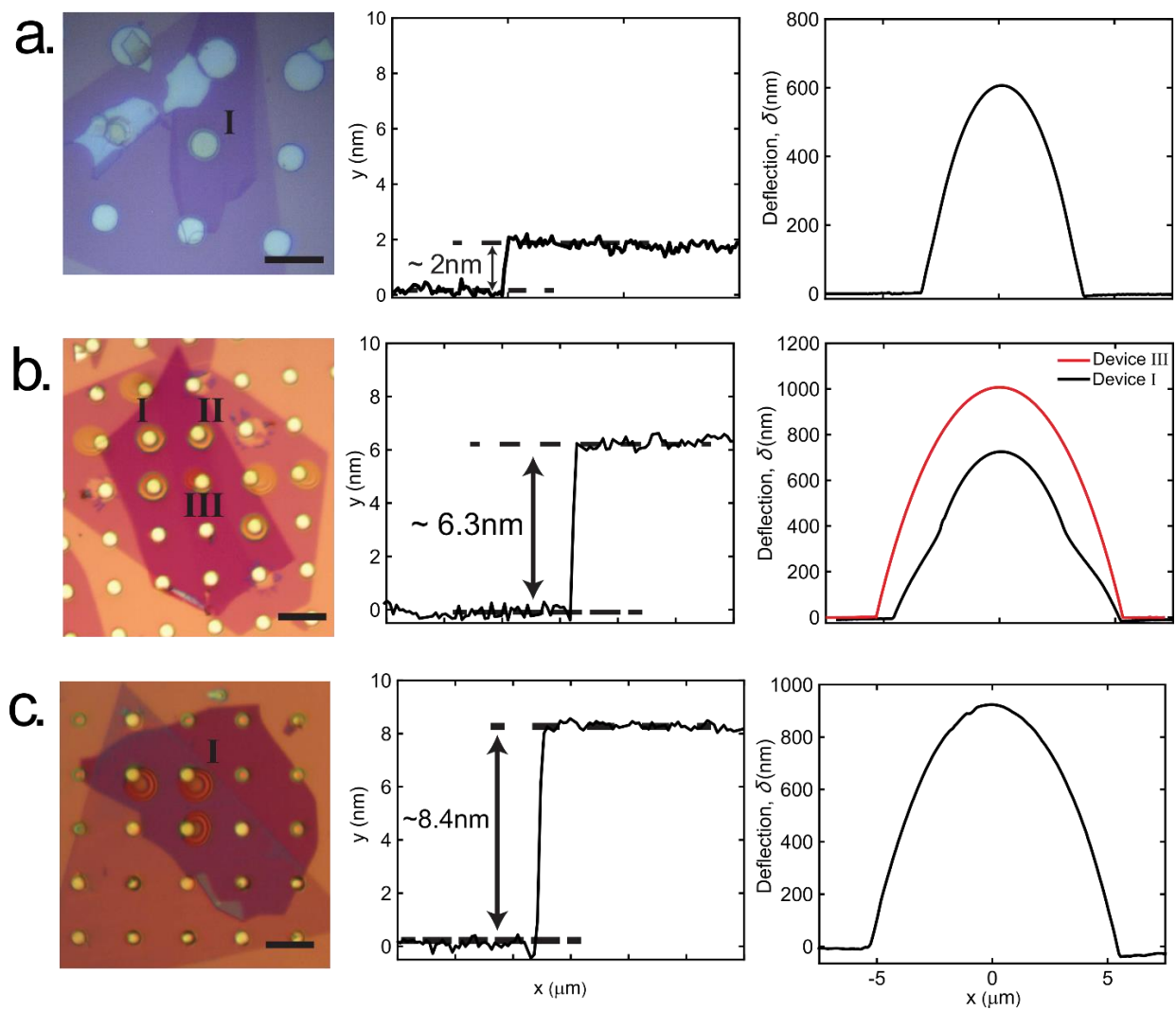

**Figure 4. a)** Optical image (left) of the 2-nm-thick device that shows LS delamination. The thickness of the LS (middle) and AFM cross-section(right) of the device(I). LS delamination is hard to observe with the AFM scan. The scale bar is 10 μm. **b)** Optical image (left) of one of the LS devices. Blister-I is typical LS delamination from $SiO_x$, Blister-II is an LS delamination from $SiO_x$, and Blister-III is the regular delamination of $MoS_2$ from the FLG surface. The thickness of the LS (middle) and cross-section of the Blister-I and III (right). The scale bar is 20 μm. **c)** Optical image (left) of $MoS_2$ on slightly thicker multilayer graphene. After a certain thickness, we only observe regular delamination of $MoS_2$ from the FLG surface. The thickness of the LS (middle) and cross-section of the Blister-I (right). The scale bar is 20 μm.

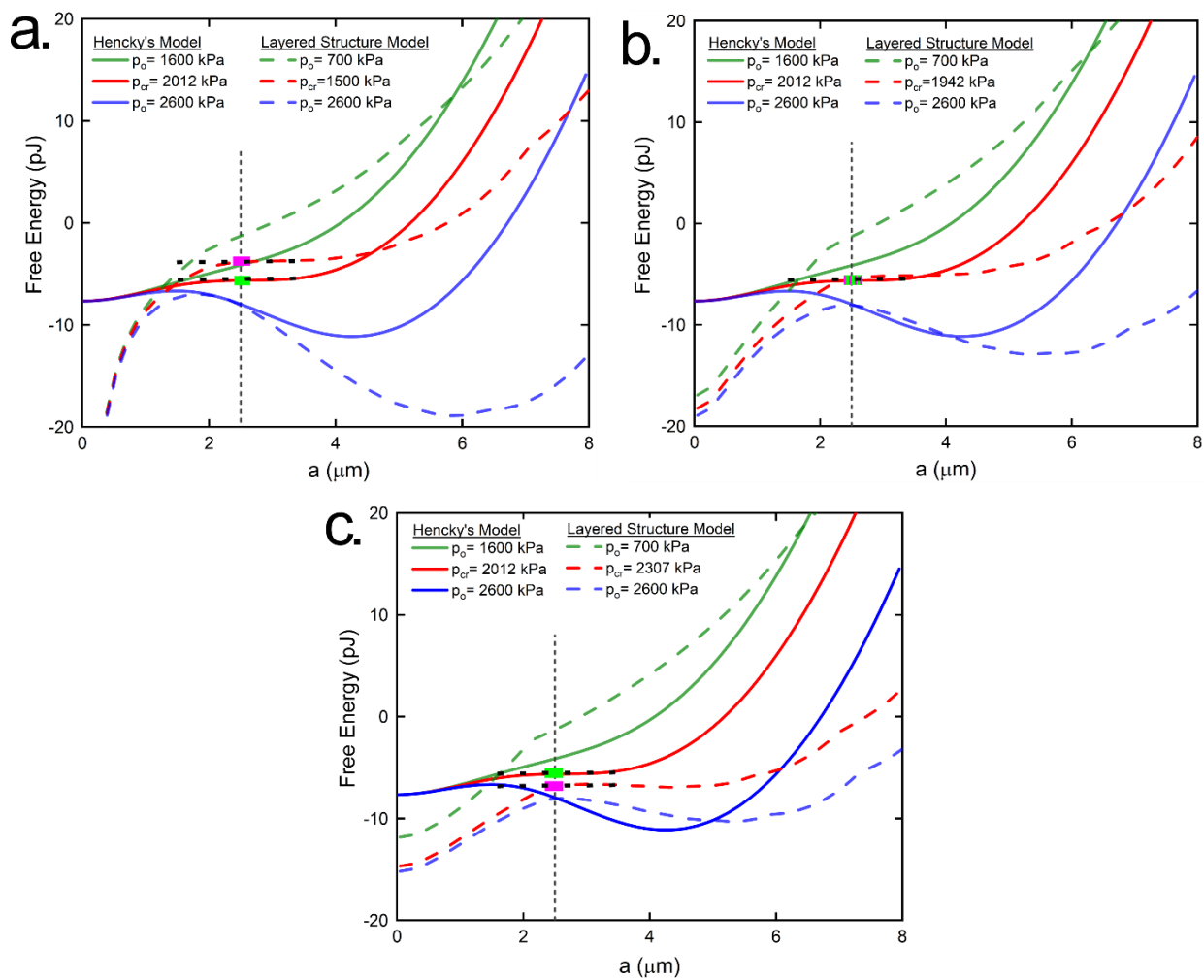

**Figure 5.** Comparison of free energy models at 3 different $p_o$. The black vertical dashed line shows the well radius. Rectangular symbols indicate the equilibrium configuration where $dF/da = 0$. MoS$_2$/FLG thickness **a)** 2 nm, **b)** 6 nm, and **c)** 9 nm.

Supporting Information

# Blister Test to Measure the Out-of-Plane Shear Modulus of Few-Layer Graphene

Metehan Calis[1], Narasimha Boddeti[2], and J. Scott Bunch[1,3*]

[1]Boston University, Department of Mechanical Engineering, Boston, MA 02215 USA

[2]Washington State University, School of Mechanical and Materials Engineering, Pullman, WA 99163 USA

[3]Boston University, Division of Materials Science and Engineering, Brookline, MA 02446 USA

*e-mail: bunch@bu.edu

## 1. Growth and Characterization

We grow monolayer $MoS_2$ flakes using chemical vapor deposition (CVD). First, a $SiO_x$ wafer is cleaned with acetone, isopropyl alcohol (IPA), and deionized water (DI) and exposed to ultraviolet (UV) light for 5 minutes. $MoS_2$ powder (Thermo Fisher Scientific, Molybdenum (IV) sulfide, 98%) is positioned in the middle of a furnace and a $SiO_x$ wafer is placed into a cooler region downstream (~ 650 – 700 °C). The furnace is purged with argon (200 sccm) to remove air quickly and make the furnace an argon-rich environment. Then the furnace is placed under vacuum and 60 sccm Ar, 0.06 sccm $O_2$, and 1.8 sccm $H_2$ are introduced inside the tube. The growth process consists of three steps: (i) heating up to 900 °C for 15 minutes, (ii) holding at 900 °C for 15 minutes, and (iii) cooling the furnace to room temperature.

Using optical contrast[1], Raman spectroscopy and photoluminescence (PL) spectroscopy, we identify monolayer $MoS_2$ flakes. Both Raman and PL spectroscopy were conducted in a Renishaw Raman InVia microscope using a 532 nm laser beam with 1200 l/mm gratings. The in-plane ($E^1_{2g}$) and out-of-plane ($A_{1g}$) vibrations located at 385 $cm^{-1}$ and 405 $cm^{-1}$ respectively[2] in Raman spectrum, and A exciton peak[3] in the PL spectrum located at 1.88 eV show that the membranes are single layered.

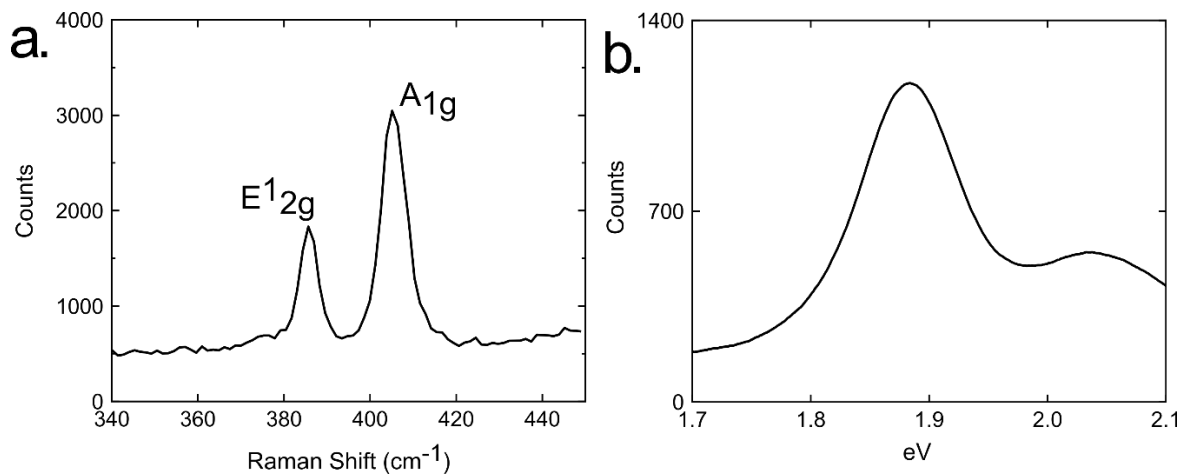

**Figure S1. a)** Raman spectrum. **b)** PL spectrum.

## 2. Graphite Substrate Fabrication and MoS$_2$ layers transfer

We start with cleaning the SiO$_x$ surface with acetone, IPA, and DI water. Utilizing the 'Scotch-Tape' method, we exfoliate the graphite flakes onto the SiO$_x$ surface by peeling off very slowly (~ 1 mm/min). Then, the graphite-covered substrate is spin-coated with photoresist S1818 at 2500 rpm and kept on the hot plate at 115 °C for 1 minute. We pattern 5 μm diameter circles onto the spin-coated chips by exposing them to UV light for 20 seconds at 8 mW power. MF-319 developer is used to remove the exposed photoresist. Reactive-Ion-Etching (RIE), with parameters 3.1 sccm O$_2$ and 25 sccm CF$_4$ at 100 mTorr pressure and 150 W power for 10 minutes is used to etch through graphite, SiO$_x$, and Si and create wells with a depth between ~480 – 700 nm. To remove the photoresist top surface, we keep the prepared chips in a bath of 1165 Remover for 12 hours at 110 °C, followed by exposure to O$_2$ plasma to remove any remaining photoresist (Fig. S2).

Before MoS$_2$ transfer, we locate few-layer graphene (FLG) wells using optical contrast and AFM scans. Subsequently, we begin spin coating the CVD-grown MoS$_2$ flakes with PMMA at 1800 rpm. We create a window on thermal-release tape, and it is stamped onto the PMMA-covered MoS$_2$ substrate. We place the PMMA-covered MoS$_2$ substrate into DI water and let the water separate the MoS$_2$ flakes from the SiO$_x$ resulting in an MoS$_2$/PMMA/thermal-release-tape (MPT) combination. We use a custom-made apparatus, to transfer the MPT combination over the etched wells, which has two main parts: a heating stage which helps adhere MoS$_2$ onto FLG wells, and a micro-manipulatable lever which allow steady approach to the FLG surface while lowering the MPT to adhere. The whole transfer process is carried out under an optical microscope and a heated (~ 85 °C) stage. With the help of the heat, the thermal-release-tape is peeled off easily and the MoS$_2$/PMMA sticks to the graphite wells. To ensure there is no trapped air inside the microcavities (resulting from the transfer process), we place them in a desiccator for a period of 12 hours. This step is carried out prior to annealing in order to prevent any potential damage that could occur during the annealing process, which takes place under vacuum condition. To remove the PMMA from the surface, we anneal the device for 7 hours at 350 °C under 20 sccm of argon flow.

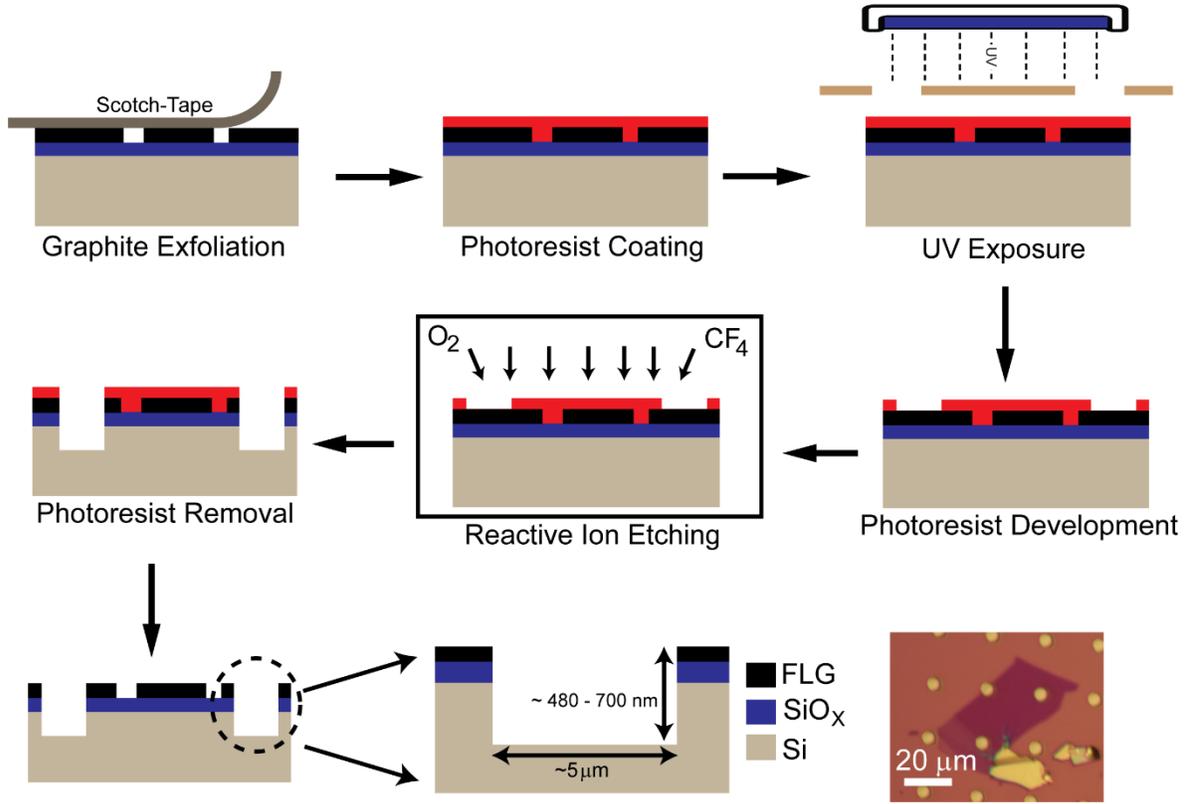

**Figure S2.** Microfabrication of the few-layer graphene wells.

## 3. Young's Modulus Calculation

To determine the Young's modulus of MoS$_2$, we use Hencky's solution [4–6] for the deformation of a pressurized clamped axisymmetric membrane, which relates the pressure difference ($p$) to deflection ($\delta$) and radius ($a$) through the formula:

$$p = \frac{K(v) E_{2D}\, \delta^3}{a^4} \tag{S3.1}$$

where $E_{2D}$ is the two-dimensional Young's modulus, $a$ is the cavity radius, and $K(v)$ is a constant which depends on the Poisson's ratio (for MoS$_2$, we use $K(v = 0.29) = 3.54$).[7] From Hencky's solution, the volume under the bulge is determined by the formula $V_b = C(v)\pi a^2 \delta$ where (for MoS$_2$, we use $C(v = 0.29) = 0.552$) and the ideal gas law ($p_0 V_0 = p_{int}(V_0 + V_b)$) is employed to determine $p_{int}$. For each input pressure, we measure $\delta$ and $a$ of each bulge with the AFM. Using Eqn. S3.1, a linear fit to the data is used to determine $E_{2D}$ for each device (Fig. S3a). The values of $E_{2D}$ are consistent with monolayer MoS$_2$ (Fig. S3b). The variation in the $E_{2D}$ values observed in the CVD-grown MoS$_2$ membranes can be attributed to sulfur vacancies[8,9], varying densities of defects[10] resulting from different growth conditions, and introducing strain[11] over the MoS$_2$ membrane due to the transfer process.

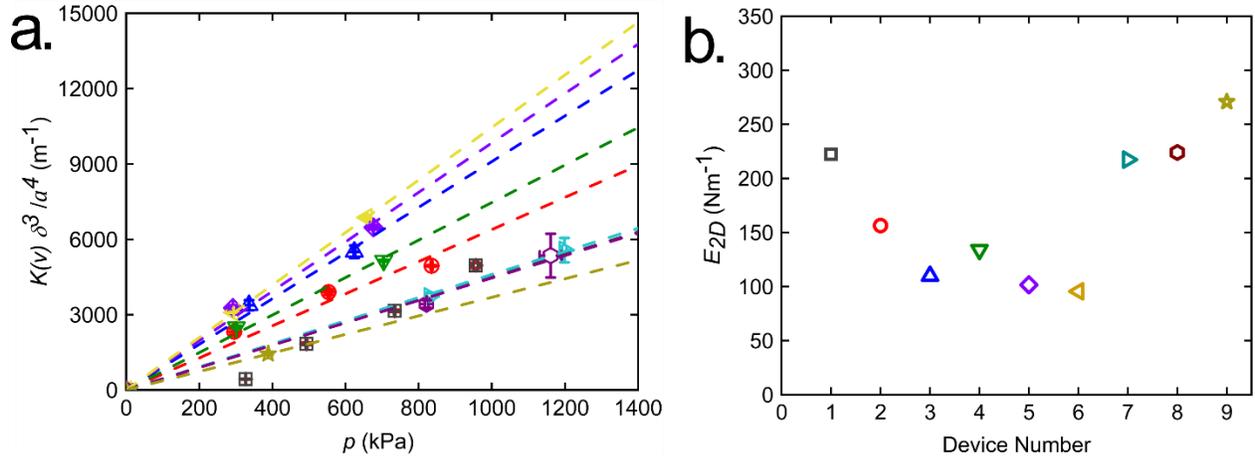

**Figure S3. a)** $K(v)\delta^3/a^4$ vs $p$ for CVD-grown MoS$_2$ membranes. Dashed lines are the linear fits used to determine the $E_{2D}$ of each device. **b)** $E_{2D}$ for each device in (a).

## 4. Shear Modulus Derivation and Free Energy Model

To compute the shear modulus of the FLG, we utilize the governing equations developed by Williams[12] for an axisymmetric membrane, incorporating a shear stress term. The governing equations derived from a force balance in the radial (Figure S4) and transverse directions for this system are:

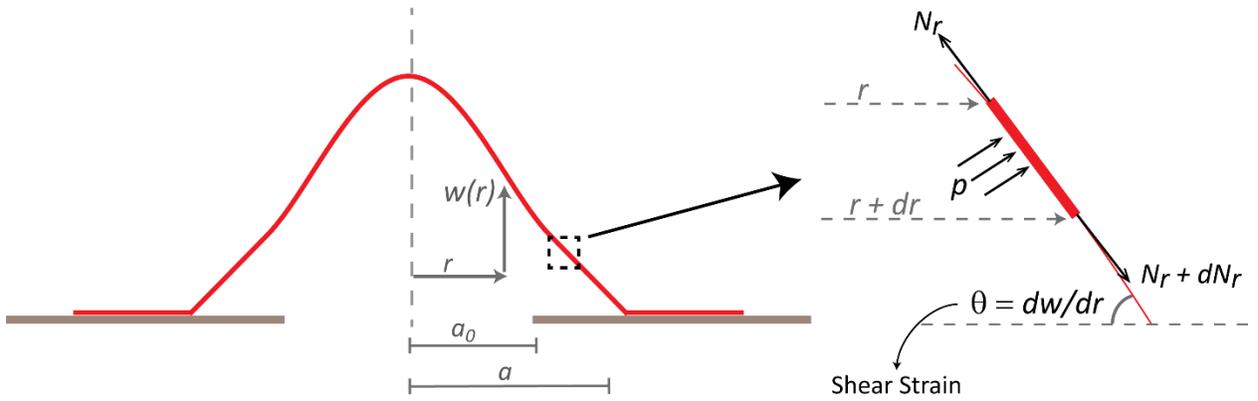

**Figure S4.** Schematic diagram of the force balance in the radial direction.

$$\frac{d}{dr}(r\, N_r) = N_t \tag{S4.1a}$$

$$\frac{d}{dr}\left(r(N_r + G_{2D})\frac{dw}{dr}\right) = -pr \tag{S4.1b}$$

$$\varepsilon_r = \frac{du}{dr} + \frac{1}{2}\left(\frac{dw}{dr}\right)^2 \tag{S4.2a}$$

$$\varepsilon_t = \frac{u}{r} \tag{S4.2b}$$

$$Et\varepsilon_r = N_r - \nu N_t \tag{S4.3a}$$

$$Et\varepsilon_t = N_t - \nu N_r \tag{S4.3b}$$

where $r$ is the radial coordinate, $u$ is radial displacement, $w$ is transverse deflection, $p$ is the pressure difference, and $E$, $t$, $v$ are the bulk Young's modulus, thickness, and Poisson's ratio of the membrane, respectively. $N_r$ and $\varepsilon_r$ are the radial stress and strain, and $N_t$ and $\varepsilon_t$ are the circumferential stress and strain. $G_{2D}$ is the two-dimensional shear modulus. $G_{2D}$ is equal to the bulk shear modulus multiplied by the thickness of the layered structure (LS) ($G * (LS\ thickness) = G_{2D}$ (N/m)). Equations S4.1a, b, S4.2a, b, and S4.3a, b can be combined to give:

$$(N_r + G_{2D})^2 \frac{d}{dr}\left(r^3 \frac{dN_r}{dr}\right) = -\frac{Etp^2}{8}r^3 \tag{S4.4}$$

This non-linear equation has been solved using a series approximation for the non-dimensional stress $f = (Etp^2a^2/64)^{-1/3} (N_r+G_{2D})$. As a result, Eqn. S4.4 can be non-dimensionalized as ($\zeta = r/a$):

$$f^2 \frac{d}{d\zeta}\left(\zeta^3 \frac{df}{d\zeta}\right) = -8\zeta^3 \tag{S4.5}$$

Furthermore, we sub-divide the whole blister into two regions along the radial direction: (i) Region I ($r \leq a_0$) is where only MoS$_2$ is suspended and (ii) Region II ($a_0 < r \leq a$) is area outside the microcavity which forms the MoS$_2$/FLG LS. Therefore, the non-dimensional stress can be written in series forms as:

$$f_i = \sum_{l=0,1,\ldots} A_{i(2l)}\zeta^{2l} = \begin{cases} Region\ I & (r \leq a_0) & i = 1 \\ Region\ II & (a_0 < r \leq a) & i = 2 \end{cases} \tag{S4.6}$$

We substitute Eqn. S4.6 into Eqn. S4.5 and equating terms on both sides we obtain $A_{i2} = -1/(A_{i0})^2$, $A_{i4} = -2/(3A_{i0}^5)$, $A_{i6} = -13/(18A_{i0}^8)$, $A_{i8} = -17/(18A_{i0}^{11})$, $A_{i10} = -37/(27A_{i0}^{14})$, $A_{i12} = -1205/(567A_{i0}^{17})$, … etc.

For the deflection profile $w(r)$ of the delaminated LS device, we carry out the same regional approach which results in:

$$w(r) = \begin{cases} w_1(r) = \left(\frac{pa^4}{Et}\right)^{\frac{1}{3}} \sum_{j=0,1,\ldots} C_j \zeta^{2j}, & \text{Region I } (r \leq a_0) \\ w_2(r) = \left(\frac{pa^4}{Et}\right)^{\frac{1}{3}} \sum_{j=0,1,\ldots} B_j \left(1 - \zeta^{(2j+2)}\right), & \text{Region II } (a_0 < r \leq a) \end{cases} \quad (S4.7)$$

First, we obtain $C_j$ and $B_j$ by equating terms with the same exponent of $r$ on both sides of the non-dimensionalized Eqn. S4.1b while using the series expression for the non-dimensional stress function, $f$ as defined in Eqn. S4.6. As a result, $C_j$ is written in terms of $A_{10}$ such that $C_2 = -1/A_{10}$, $C_4 = -1/(2A_{10}^4)$, $C_6 = -5/(9A_{10}^7)$, $C_8 = -55/(72A_{10}^{10})$, $C_{10} = -7/(6A_{10}^{13})$, …, etc., and $B_i$ in terms of $A_{20}$ such that $B_0 = 1/A_{20}$, $B_2 = 1/(2A_{20}^4)$, $B_4 = 5/(9A_{20}^7)$, $B_6 = 55/(72A_{20}^{10})$, $B_8 = 7/(6A_{20}^{13})$, $B_{10} = 205/(108A_{20}^{16})$, …, etc. Next, we determine $A_{20}$ by utilizing the fixed boundary condition $u(r=a) = 0$ for region II which results in:

$$\frac{d}{d\zeta}(\zeta f_2) - v f_2 = (1-v) f_0 \quad (S4.8)$$

where $f_0 = 4G_{2D}/(Etp^2a^2)^{1/3}$. Subsequently, $A_{10}$ can be obtained by using the governing equation and continuity condition $u(r = a_{0-}) = u(r = a_{0+})$ which results in:

$$\frac{d}{d\zeta}(\zeta f_2) - v f_2 - (1-v) f_0 = \frac{d}{d\zeta}(\zeta f_1) - v f_1 \quad (S4.9)$$

Finally, we obtain the remaining unknown $C_0$ by enforcing the continuity condition $w^-(\zeta = a_0/a) = w^+(\zeta = a_0/a)$.

As the next step, we model the blister as a thermodynamic system. This involves developing a free energy model that comprises four parts.

$$F = F_{mem} + F_{gas} + F_{ext} + F_{adh} \quad (S4.10)$$

$F_{mem}$ refers to the strain energy term resulting from the stretching caused by the application of a pressure load on the membrane. It is expressed as:

$$F_{mem} = \int_0^a \left(\frac{1}{2}(N_r\epsilon_r + N_t\epsilon_t)\right) 2\pi r dr + \int_{a_0}^a \frac{1}{2} G_{2D} \left(\frac{dw}{dr}\right)^2 2\pi r dr \tag{S4.11}$$

We divide the integral of the MoS$_2$ membrane's strain energy contribution due to the stretching (first term in Eqn. S4.11) into two regions and introduce the parameter $\rho = a_0/a$ to non-dimensionalize the expression. We define $U_{strain1}$ is the non-dimensionalized strain energy within the well region ($\rho < a_0/a$), and $U_{strain2}$ is that within the delaminated region of the blister ($a_0/a < \rho < 1$). This results in:

$$U_{Strain1} = \int_0^\rho \left((f_1)^2 + \frac{d}{d\zeta}(\zeta f_1)^2 \right) \zeta \, d\zeta - v \int_0^\rho \left(2f_1 \frac{d}{d\zeta}(\zeta f_1)\right) \zeta \, d\zeta \tag{S4.12}$$

$$U_{Strain2} = \int_\rho^1 \left((f_2 - f_0)^2 + \frac{d}{d\zeta}(\zeta(f_2 - f_0))^2 \right) \zeta \, d\zeta - v \int_\rho^1 \left(2(f_2 - f_0)\frac{d}{d\zeta}(\zeta(f_2 - f_0))\right) \zeta \, d\zeta \tag{S4.13}$$

Next, we proceed to evaluate the second term in $F_{mem}$, which accounts for the shear contribution. We express this term as in its non-dimensionalized form as ($\hat{w} = w \left(\frac{pa^4}{Et}\right)^{-1/3}$ non-dimensionalized deflection):

$$U_{Shear} = \int_\rho^1 \zeta^2 \left(\frac{d\hat{w}}{d\zeta}\right)^2 \pi d\zeta \tag{S4.14}$$

and the final expression for $F_{mem}$ is,

$$F_{mem} = \left(\frac{pa^4}{Et}\right)^{1/3} \left(\frac{\pi p a^2}{16}(U_{Strain1} + U_{Strain2}) + \left(\frac{pa^4}{Et}\right)^{1/3} G_{2D} U_{Shear}\right) \tag{S4.15}$$

$F_{gas}$ is the energy change due the expansion of the gas molecules trapped in the blister,

$$F_{gas} = -p_0 V_0 \ln\left(\frac{V_0 + V_b}{V_0}\right) \tag{S4.16}$$

$F_{ext}$ is the energy change of the external environment,

$$F_{ext} = -p_{ext} V_b \tag{S4.17}$$

$F_{adh}$ is the adhesion energy of the LS-substrate interface,

$$F_{adh} = \Gamma \pi (a^2 - a_0^2) \tag{S4.18}$$

where $\Gamma$ is separation energy per unit area. During the analysis of the free energy, we utilize multiple terms (up to 5 terms) in the non-dimensionalized stress function, $f$. However, as we add more terms to the series, the computational cost increases. After comparing the results, we observed that beyond the inclusion of 3 terms in the series, there is no significant change, and the results converge to the same approximated value. Therefore, we made the decision to utilize the 3-term expansion, which yields satisfactory results in numerical solutions. With this approach, we plot the input pressure vs. shear modulus for all tested devices in Figure S5. By conducting the blister test at various input pressures, we induce multiple instances of delamination in the devices.

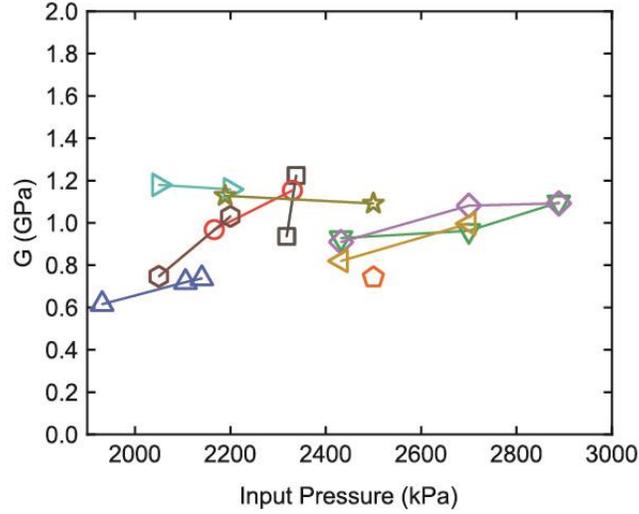

**Figure S5.** Shear modulus development of each device against varying input pressure. 9 devices undergo multiple delamination from the surface.

## 5. Investigation of the Delamination Behavior of the LS System by Varying the Thickness

In the experiments, we observed two distinct delamination behaviors across different devices: (i) only the MoS$_2$ membrane delaminates from the surface of FLG, and (ii) MoS$_2$/FLG LS delamination from the SiO$_x$ surface. The primary distinction between the devices lies in the varying thickness of the FLG. To comprehend this transition, in Figure S6, we plot the critical pressure ($p_{cr}$) at which the blister first delaminates vs. the thickness of the LS. Employing the microcavity dimensions (well depth = 600 nm, well radius = 2.5 μm), we calculate the critical pressure ($p_{cr\text{-}Hencky}$), such that only the MoS$_2$ membrane separating from the FLG, using Hencky's free energy model.[4,13] To calculate the critical pressure, we use the expression:[14]

$$\Gamma_{sep} = \frac{5}{4} C K E_{2D} \left(\frac{\delta}{a}\right)^4 \tag{S5.1}$$

First, by rearranging the Eqn. S5.1, we determine the critical deflection just before the delamination occurs ($\Gamma_{sep} = 0.39$ J/m$^2$, $C = 0.522$, $K=3.55$, $E_{2D}= 180$ N/m). Next, we calculate bulge volume ($V_b$) and obtain $p_{int}$ by using Eqn. S3.1. Finally, we substitute these values back into ideal gas law to find the $p_{cr-Hencky}$. Additionally, we plot the critical pressure ($p_{cr-LS}$) for MoS$_2$/FLG LS delamination from the SiO$_x$ surface in Figure S6. To do this, we numerically solve the LS model we developed by incorporating the calculated mean $G$ value ($G = 0.97$ GPa) as well as values, $G = 1.12$ and $0.82$ GPa, that are one standard deviation from the calculated mean $G$ value. This helps us establish the range within which $p_{cr-LS}$ corresponds to the pressure just before the diameter expansion occurs. See main text for the parameters used in the calculation. Based on the intersection of $p_{cr-Hencky}$ and $p_{cr-LS}$ we define 2 zones: (i) Zone #1 includes the region up to the intersection of $p_{cr-LS}$ ($G = 0.82$ GPa) and $p_{cr-Hencky}$ where we observe delamination of LS, (ii) Zone #2 encompasses the region starting from intersection of the $p_{cr-LS}$ ($G = 1.12$ GPa) and $p_{cr-Hencky}$. The hatched area is what we call the transition zone, where we observe both LS (MoS$_2$/FLG) and regular (only MoS$_2$) delamination. In this region, the free energy curves (see the main text) from the LS model and Hencky's model reach their equilibrium configuration ($dF/da = 0$) at very similar critical pressures. Thus, small variations in adhesion strength, and surface–LS interaction can lead to either regular or LS delamination when $p_o > p_{cr}$. The intersection point of $p_{cr-LS}$ ($G = 0.82$ GPa (red line)) and $p_{cr-Hencky}$ (dashed line) provides the cutoff thickness below which we expect LS delamination is preferred over MoS$_2$ separation from FLG. We also populated Figure S6 with two sets of data points: (i) black data points represent devices that exhibit FLG/MoS$_2$ LS delamination from the SiO$_x$ surface, and (ii) purple data points show regular delamination where only the monolayer MoS$_2$ membrane separates from the FLG surface.

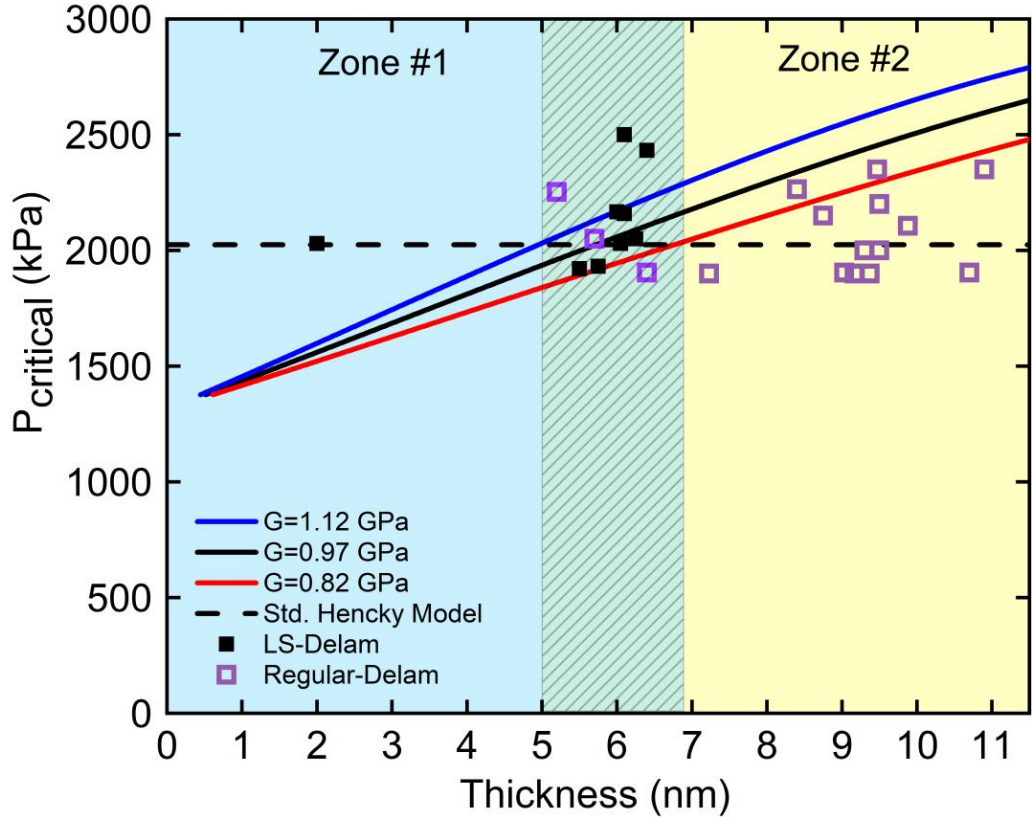

**Figure S6.** $p_{cr}$ vs thickness of MoS$_2$/FLG LS device. The solid lines show the expected $p_{cr}$ using the measured average shear modulus (black), plus (blue), and minus (red) one standard deviation. The dashed line is the expected $p_{cr}$ for delamination of MoS$_2$ from the FLG surface using the standard Hencky's solution. The plot is divided into two zones; (i) Zone #1 where we observe only LS delamination, (ii) Zone #2 where we observe only MoS$_2$ delamination, and the hatched transition zone where we both.

## 6. Raman Spectroscopy Analysis over the Delaminated Blister

We performed Raman spectroscopy to verify the delamination of graphite flakes with the MoS$_2$ membrane. We carry out the scan through points located along the dashed line shown in Figure S7a and sub-divide the line scan into 5 regions. Region 1 (Fig. S7b) and region 5 (Fig. S7f) are on the supported area where there is no delamination. In region 2 (Fig. S7c) and region 4 (Fig. S7e), we observe LS delamination where both FLG and MoS$_2$ separated from the surface together and we measure the graphite G and 2D Raman peaks[15], however the intensities are lower compared to the supported area presumably due to interference.[16] In region 3 (Fig. S7d), we have only the MoS$_2$ membrane suspended over the microcavity. Signals of the 2D peak from FLG disappear and the MoS$_2$ Raman resonance peaks become more prominent.

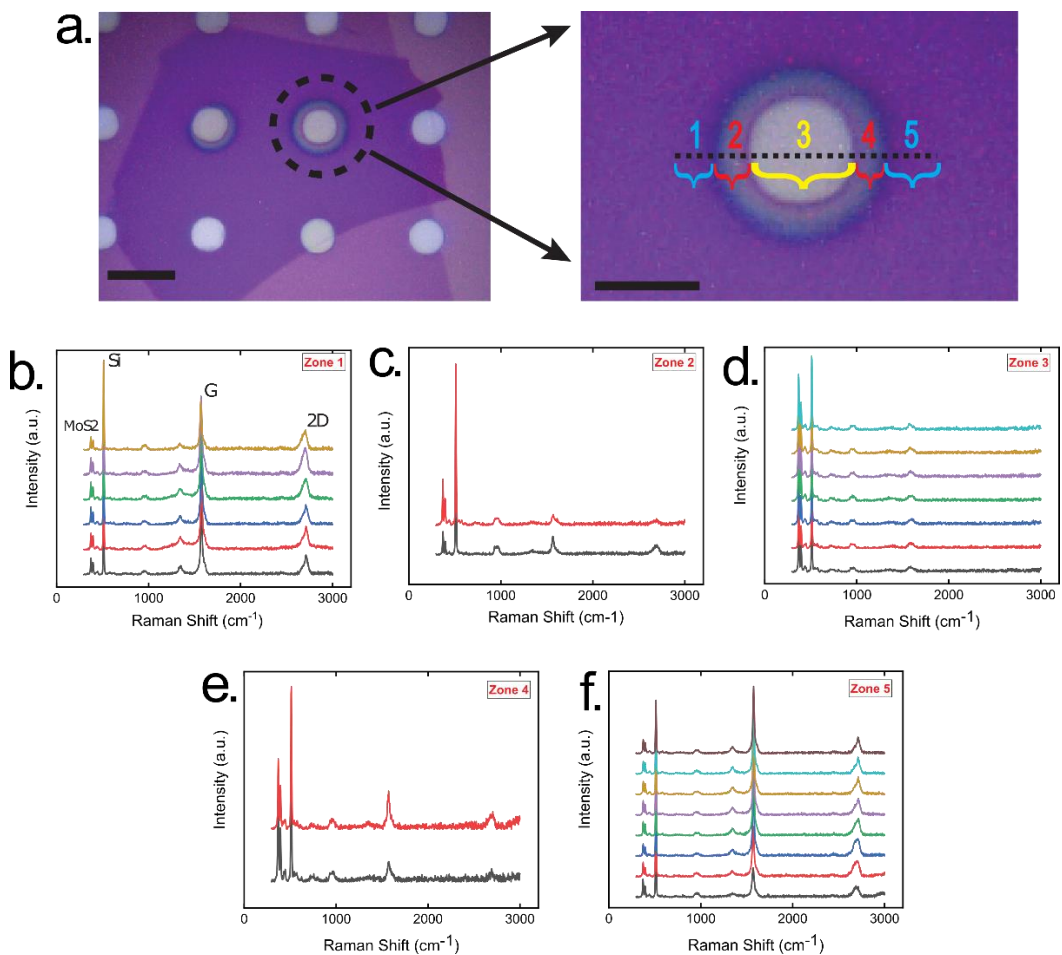

**Figure S7. a)** Optical image of FLG/MoS$_2$ which shows LS delamination labeled with a circled dash line. The scale bar is 10 μm (Left image). Raman spectroscopy is carried out over points which lie on the dashed line (Right image – Scale bar is 5 μm) We sub-divided Raman spectrum into 5 regions. **b)** and **f)** are Raman spectrum over the supported area with no delamination. In (b), we labeled the peaks correspond to few-layer graphene (G and 2D), Si, and MoS$_2$. **c)** and **e)** are Raman spectrum over the FLG/MoS$_2$ with LS delamination. **d)** Raman spectrum over the microcavity where only MoS$_2$ is suspended.

## 7. Estimation of Bending Strain Energy

As the thickness of the graphite increases in the LS devices, bending strain energy also needs to be considered in the free energy calculation. To ascertain the necessity of this addition, we begin by determining the bending modulus applicable to our tested devices. Since the LS is primarily composed of few-layer graphene, we used the bending rigidity of graphene in our bending strain energy estimates and used the assumptions of classical plate theory where:

$$D = \frac{E h^3}{12 (1 - v^2)} \tag{S7.1}$$

where $E$ is Young's modulus, $h$ is the thickness and $v$ is the Poisson's ratio.

For a 6nm thick (~ 17 layers of graphene) LS case, we calculate a bending rigidity of 1.97 x 10$^{-14}$ J.

The strain energy contribution due to bending by exploiting the expression formularized by Timoshenko[17,18] yields:

$$U_{bending} = D\pi \int_a^{a_0} \left[ \left(\frac{d^2w}{dr^2} + \frac{1}{r}\frac{dw}{dr}\right)^2 - 2(1-v)\left(\frac{d^2w}{dr^2}\frac{1}{r}\frac{dw}{dr}\right) \right] r d\theta dr \tag{S7.2}$$

For deflection $w(r)$, we approximate it by using the expression:

$$w(r) \approx \frac{a^2 - r^2}{2R} \tag{S7.3}$$

where $R$ is the radius of the curvature. As a result, the bending strain energy estimate is:

$$U_{bending} = \frac{\pi D (1 + v)}{R^2}(a^2 - a_0^2) \tag{S7.4}$$

From equation S7.4, we calculate and tabulate results of the bending strain energy and shear strain energy for the 3 tested devices in Table S1.

|  | Input Pressure($p_0$) (kPa) | Shear Strain Energy (J) | Bending Strain Energy (J) | Thickness (nm) |
|---|---|---|---|---|
| **Device ID: Q_W23** | 2318 | 2.80 x 10$^{-12}$ | 2.50 x 10$^{-15}$ | 6.25 |
| **Device ID: C2_S18** | 2166 | 1.71 x 10$^{-12}$ | 1.83 x 10$^{-15}$ | 6 |
| **Device ID: AH_X29** | 2106 | 3.98 x 10$^{-12}$ | 3.19 x 10$^{-15}$ | 5.75 |

**Table S1.** Comparison of calculated strain energy contributions for 3 tested devices.

It can be seen that bending strain energy is 3 orders of magnitude less than the shear strain energy in this thickness range. Therefore, we omit the contribution of bending from the free energy model.

## 8. Thickness Dependence of the LS Delamination Profile

In Figure S6, it can be observed that the devices which show LS delamination mostly lie within the 5 nm to 7 nm thickness range. The main reason is that when the combined thickness (thickness of MoS$_2$ + thickness of few-layer graphene) is below 5 nm, the LS delamination becomes hard to distinguish from a regular delamination in the AFM profile. To illustrate this, using the same device parameters (radius and depth) and $G = 0.97$ GPa, we plot the delaminated profile of the blisters at critical pressure as a function of varying device thickness in Figure S8. When the FLG thickness is zero ($f_0 = 0$), the result of the LS model matches that of Hencky's free energy model (dashed line in Figure S8).

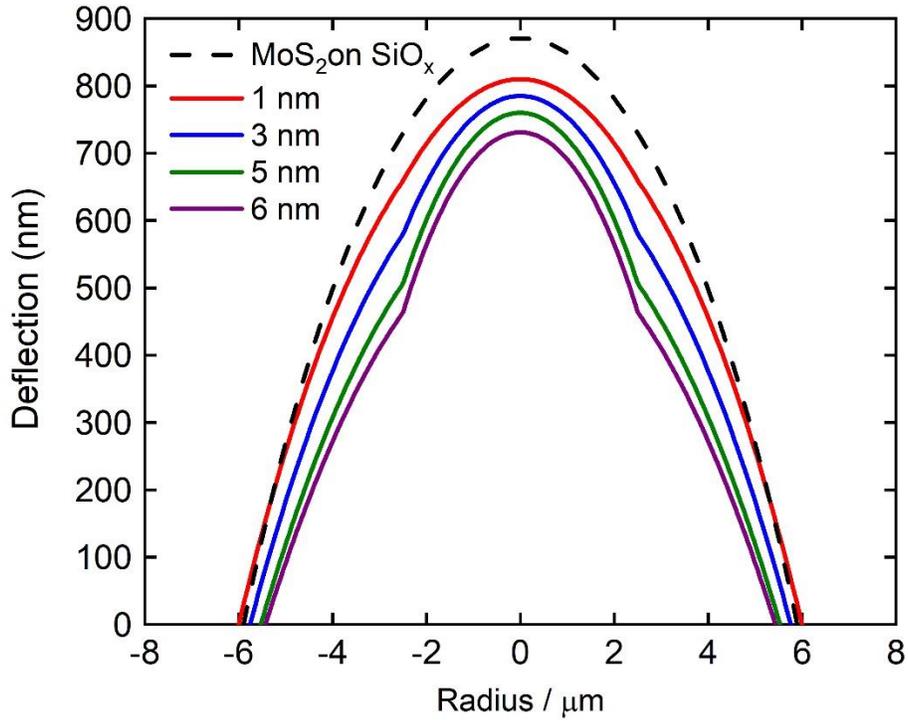

**Figure S8.** Profile change correspondences to varying combined thickness of LS.

## 9. Non-Circular Delamination Example

In our experiments, we also observe asymmetric/non-axisymmetric delamination of $MoS_2$/FLG LS from the $SiO_x$ surface. In Figure S9 and S10, we show AFM amplitude images from two different devices. Before delamination, the $MoS_2$ bulges stay axisymmetric, and at higher pressures, the $MoS_2$/FLG begins to delaminate from the surface. The LS delamination occurs radially outward in a preferred direction. We attribute this to possible surface inhomogeneities around the blister[19], or structural defect.[20,21]

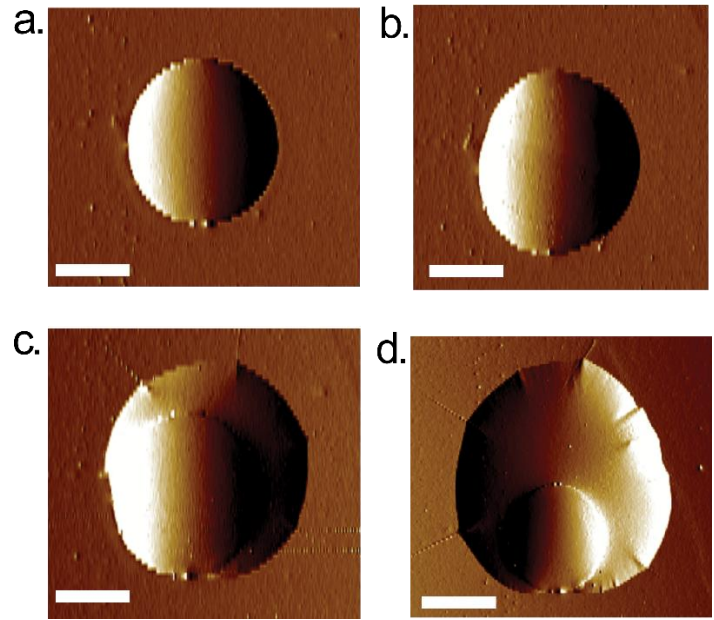

**Figure S9.** AFM amplitude image of device #1 with non-circular delamination. **a)** at input pressure ($p_0$) = 1609 kPa **b)** at $p_0$ =1780 kPa **c)** at $p_0$ = 2325 kPa **d)** at $p_0$ = 3634 kPa. Scale bars are 2.5 µm.

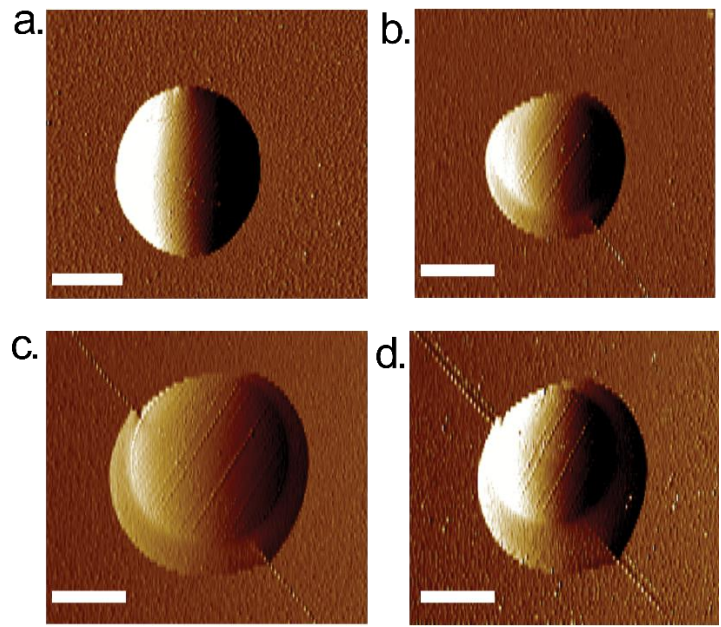

**Figure S10.** AFM amplitude image of device #2 with non-circular delamination. **a)** at $p_0$ = 1300 kPa **b)** at $p_0$ =1588 kPa **c)** at $p_0$ =1620 kPa **d)** at $p_0$ =1906 kPa. Scale bars are 2.5 µm.

We also observe MoS$_2$/FLG undergoing dramatic delamination when two devices are in close proximity and coalescing together resulting in a very large and irregular-shaped blister (Fig. S11 and Fig. S12).

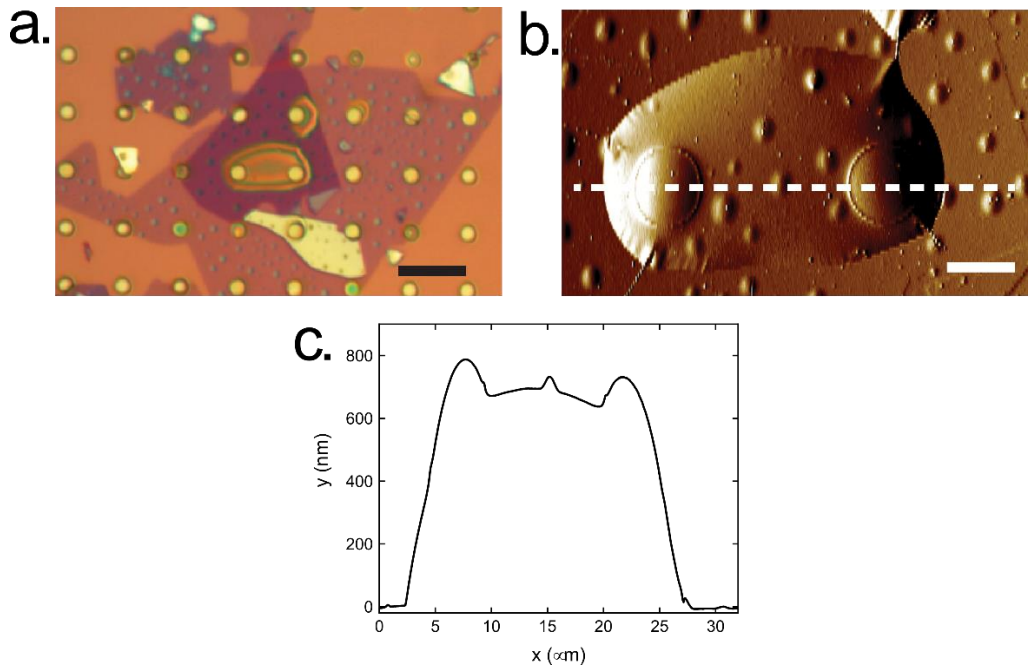

**Figure S11. a)** Optical image (Scale bar is 20 µm) and **b)** AFM amplitude image (Scale bar is 5 µm) of two blisters coalesced together. **c)** Deflection profile of the large delaminated configuration which is labeled with a dashed line in (b).

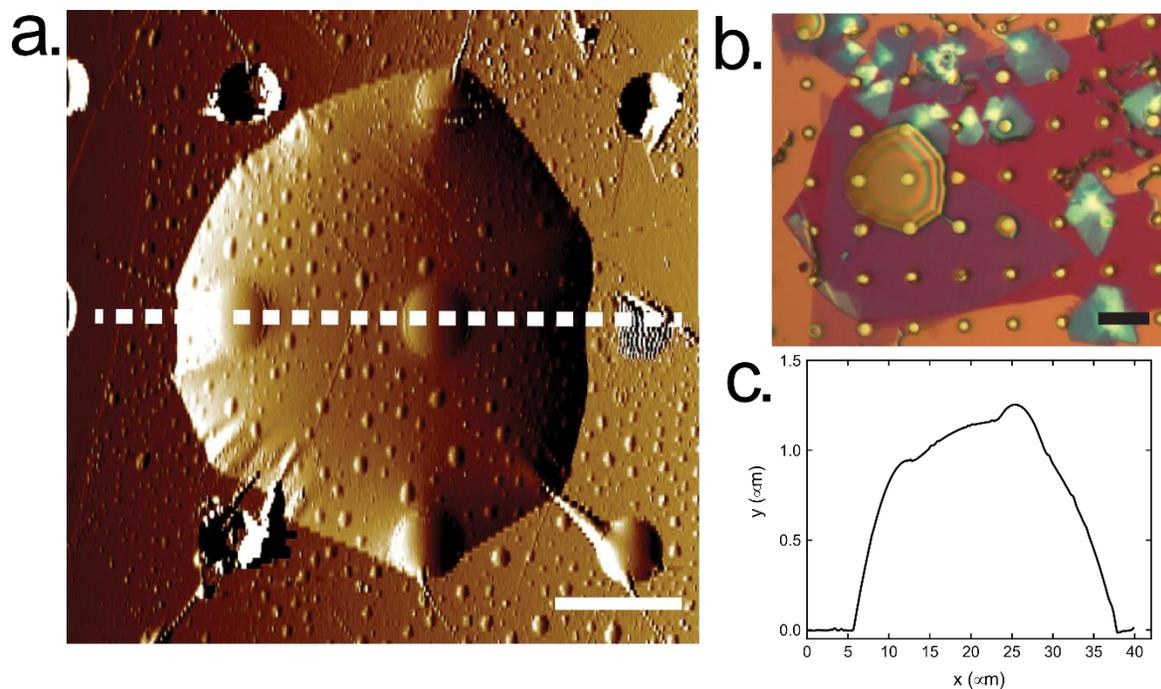

**Figure S12. a)** AFM amplitude image (Scale bar is 10 μm) and **b)** Optical image (Scale bar is 20 μm) of two blisters coalesced together. **c)** Deflection profile of the large delaminated configuration which is labeled with a dashed line in (a).